# Augmented Body: Changing Interactive Body Play

Matthew Martin
Auckland University of Technology
Private Bag 92006
Wellesley Street, Auckland, NZ.
+64 (0)9 921 9999
mamartin@aut.ac.nz

James Charlton
Auckland University of Technology
Private Bag 92006
Wellesley Street, Auckland, NZ.
+64 (0)9 921 9999
james.charlton@aut.ac.nz

Andy M. Connor
Auckland University of Technology
Private Bag 92006
Wellesley Street, Auckland, NZ.
+64 (0)9 921 9999
andrew.connor@aut.ac.nz

## ABSTRACT
This paper investigates the player's body as a system capable of unfamiliar interactive movement achieved through digital mediation in a playful environment. Body interactions in both digital and non-digital environments can be considered as a perceptually manipulative exploration of self. This implies a player may alter how they perceive their body and its operations in order to create a new playful and original experience. This paper therefore questions how player interaction can change as their perception of their body changes using augmentative technology.

## Categories and Subject Descriptors
H.1.2 [**User/Machine Systems**]: Human information processing

## 1. INTRODUCTION
As technology becomes more ubiquitous and immersive it has become possible for new 'realities' to emerge [1]. Novel concepts such as mixed reality, augmented reality, augmented virtuality and diminished reality can influence our interpretation of ourselves and the spaces in which we exist. All of these new realities merge with or replace parts of the physical world, and share common characteristics or goals. As far back as 1913, Edmund Husserl discussed how the artificial world interacts with the physical world of everyday human activities in order to enrich the experiences of perception, affordance and engagement [2]. That ideal of enrichment in terms of perception of self can inform explorative ideas through the embodying of digital objects, specifically personally accustomed physical objects, while keeping themselves within the physical realm. This paper discusses issues surrounding expected interface interaction and habits of play as well as theories of the compelling nature augmented body parts holds in today's virtual reality applications. This paper presents early outcomes of a system that facilitates body part replacement with digital objects and has the potential to support experimentation around the alteration of self-perception that occurs as a result of such replacement. This differs from much of the previous work in augmented realities that allow users to grasp and manipulate so called foreground digital objects by coupling the digital with physical objects [3]. Instead, the approach outlined in this paper focuses on replacing body part and function instead of viewing the digital body objects as something to interact with. In essence, the approach allows a user to have the digital object become part of themselves which has the potential to promote a much more playful interaction.

## 2. AUGMENTED MINDSET
Virtual reality can be used for creating worlds of imagination that allow the user to interact with objects which are not genuinely there [4]. When an individual submits themselves to a virtual reality anything can be believed to be possible. This virtual world is the forefront of perception, shielded from the gravitations of reality. No boundary or law permitted in the real world needs to be translated to a digital one, either directly or through interpretation in terms of factors of realism. The virtual reality allows a focus on environment expansion and expression, which comes hand in hand with a need for "presence". The concept of presence refers to the phenomenon of behaving and feeling as if we are in the virtual world created by the computer [5]. Immersive virtual environments can break the deep, everyday connection between where our senses tell us we are, where we are actually located and whom we are with [5]. For some people, being in a virtual world mindset may cause difficulty connecting the real body interaction while in a digital environment. This would occur when sensory data from one environment (say, kinesthetic and tactile information from the real world) and different data from a competing environment (say, visual and auditory data from a computer generated virtual environment) are simultaneously received and the virtual reality comes to dominate [6].

Keeping a person's mindset and perception within real spaces creates a different reaction for body interaction. The core focus point for the person is on what is digitally different while keeping within the parameters of reality. By limiting digital augmentation to a single body part there is a different interaction between the perception of the body part and the perception of function in the real world. This is distinctly different than many approaches to embodiment in virtual environments that focus on the representation of the whole body in a virtual world [7]. With only a section of an individual's body appearing as replaced by a digital or virtual object the player can construct their embodiment for digital interaction and movement with their body parts. One such example could be a "virtual mirror" that shows virtual objects projected onto the user [8], though in this case rather than the projection of an additional object the outcome is more of a playful virtual prosthesis.

Restricting augmentation and digital change to a person's body promotes an internal interaction development, allowing for new exploration within the real space and promoting acceptance of the digital change as part of the real world. Focusing on semi-believe and believable body alteration for a person in a real space can create new actions not possible in a fully immersive environment.

What this means is while the body of the person is mostly the same, creating smaller changes (such as a hand becoming a hook or a leg becoming a wheel) it pushes for simpler and effective body interaction only possible from smaller augmented adjustments.

## 3. PHYSICAL & DIGITAL INTERACTION

The types of interactions the body makes with computers has not progressed or change much since the 1960s [9]. The most common types of interface today such as the computer mouse, stylus and even gesture recognition can all trace their origins to this time. The computer mouse was developed in 1965 [10] and is constructed around the hand as if stating the hand as a rigid part of design and the mouse a purely external tool. This is emphasised by the original paper that instructs:

> 'To begin making the screen selection, his right hand leaves the keyboard and takes hold of ("accesses," in our terminology) the selection device. By moving this device he controls the position on the screen of an associated tracking mark (or "bug"), placing it over the "target" text entity.' [10]

Whilst more modern interface technologies exist and are growing in popularity, even touch screen and touchless gesture systems create a divide between the body and the digital object, though one that is less apparent. There have been attempts to consider more direct interactions, for example the Skinput system [11] utilises vibrations through the body as an input mechanism and then appropriates the body itself as an input surface. Similarly, various tangible and haptic interfaces are under development [12] that challenge traditional views of an interface, however whilst these all blur the division between the digital object and the body they all rely on a fixed view of the body itself.

If we think the body as not just a fixed template it opens up possibilities for new alterations for system synergy. For example, recent advances in wearable robotics demonstrate the potential for new interactions by changing what the body is possible of achieving [13]. The concept of a seven fingered hand that emerges from the work of Yu & Asada [13] creates new accessible points for interaction, however freeing the idea from the need for a physical augmentation allows greater scope for both usefulness and playfulness. For example, rather than a seven fingered hand it is possible to imagine seeing a virtual eight tentacle arm coming from the body which creates a huge potential for multiple interactions that are not currently possible in the real world. A wide range of possible examples can be envisaged, from more "serious" cases of experiencing prosthetics, through to more fun examples such as musical instruments, toys or household items such as vacuum cleaners and egg whisks. The production of such extensions to the body can easily be achieved and is therefore a plausible route for augmented body interaction alteration. It allows interaction with physical objects, manipulating body movement effectively, however the use of augmenting such an extension or more to the point a replacement, can create fluid body change as it looks to interact with the environment. Even if a digital body change is not a tangible piece it could still be a possibility for both external computer interaction and human body adjustment.

As these augmented body parts become capable of communicating with other external systems, it may be possible to eliminate the need for tangible components completely. Such external systems can be developed specifically for effective use with augmented parts of a person's body. While not a revelation, it makes for new thoughts on design around human capabilities as a consequence of digital extensions and replacements. A person does not need to be thought of as the body parts they have but the digital parts they could potentially have and how that changes their accessibility and interaction potential.

With augmented technology it is already possible to move past the tangible objects and into digital ones. A number of examples are present in the literature, for example the usage of an artificial limb or body [14, 15] or the design and fitting of prosthetic limbs in a virtual environment [16]. The latter of these examples illustrates the capability for the interaction can be with a non-real component that is not part of the fixed template of the body. These examples also have one thing in common, that they are very much focused on "serious" applications. Attributes such as 'fun' and 'pleasure' are more abstract, and there are uncertainties as to how the different possibilities for supporting playful experiences can be addressed through some form of digital prosthesis, however this uncertainty also encourages an exploration of this possibility.

With augmenting digital object parts replacing a person's perception of the body, it is important to think of its conflicts that occur with already established interactions of play with the limbs. By changing a player's hand to being thought of as something else it could cause difficulty for use with standard controllers for normal hand interaction. This in itself promotes the emergence new forms of play and discovering the new types of interaction. Many games utilise player expectation for gameplay experience purposes and manipulate habitual control patterns [17, 18]. Developing new rule breaking methods has the potential to create explorative and puzzle inducing cognitive engagement, establishing an interactive enjoyment and natural interest for people too used to traditional game movement. The use of augmenting a person's body can have similar effects and uses, allowing themselves to rediscover movements and basic communication as they feel their body has become changed.

## 4. IMPLEMENTATION

The work constructed in this research experimented with limb replacement using digital manipulation and interaction. An example of similar work is the digitally disfiguring hand artwork presented by Golan Levin [17]. It presents ways of causing uncomfortable sensations in the hand by projecting on a screen a user's hand with digitally disfigured alterations. In the work illustrated here the user places their hand between a monitor and a Leap motion controller as shown in Figure 1. The user can look onto the monitor and observe an object in the position of where their hand used to be.

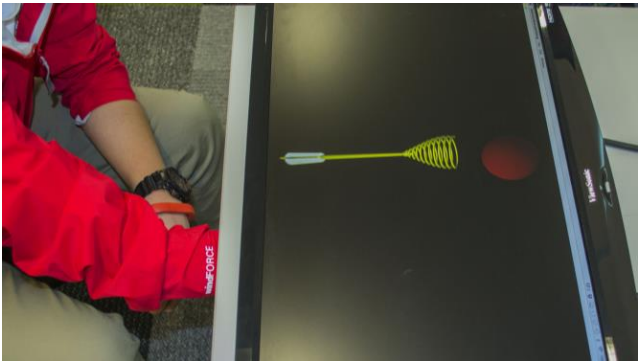

**Figure 1.** An approach to replacing user's arm with a 3D whisk model.

This is to produce a similar effect to the famous rubber hand experiment, which tricks a person into embodying a rubber hand instead of their own [18]. As the person moves and acts with their hand and arm underneath the monitor, the object displayed on screen follows in kind. The object replacing the hand can be a physical object, scanned using the a motion controller or video feed in real time as shown in Figure 2, or a 3D model of an object created digitally.

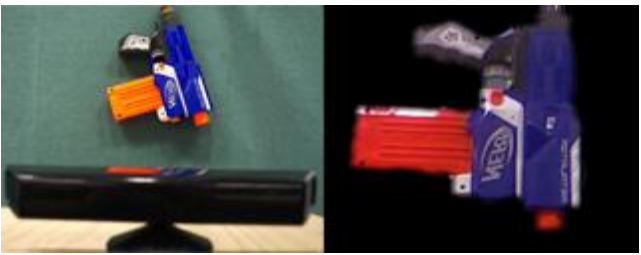

**Figure 2.** The object to replace the arm scanned with a Kinect.

The purpose of scanning physical objects was to give the user a connection to such items as being believable replacements, particularly as the object replacing the hand is one the user can instantly connect to in the physical world. The intention of replacing the body parts digitally with real objects extends the mindset by focusing on believable body alteration. This translates limitations of the real world object function to the user's believed action.

Alternatively, the benefits of using a 3D rendered model is its ability to be abstracted from hand sized or hand related objects, while having the user affiliate expected function or attributes of the object. Testing began first with hand related objects such as hammers, whisks and paintbrushes with later testing concluding on vehicles, animal limbs and insects. Interactions to perform with these hand-replacing objects included moving a virtual ball and line drawing.

The task chosen for the user was an important component to the type of behavior pattern they associated with their new body part. For instance, if a user were to try grabbing a virtual ball using a digital paw they would imitate their hand as best to their ability in similar fashion. That is to say without the use of the opposable thumb as part of that act and as a swiping action. Likewise if there were a butterfly instead of the paw, the user would feel inclined to move delicately and slowly interact with the virtual ball. As all the objects were static in nature it became apparent how lasting these types of behaviors of the user were to be and resulted in the user falling back into their basic arm movement pacing and positioning.

The line drawing tasks revealed different flows of action when equipped with different hand-based replacements. If the object were a pen the user would attempt to draw as if holding a pen. An airbrush gun would instead become a direct pointing action, using the nozzle as the trigger for drawing.

From the tests conducted it emerged how feasible it is to trick the mind about what is part of the human body. While the experiment gave a sense of embodying an object and replacing the physical arm, it was limited to the space of interaction and environmental impact. However, despite this limitation the digital prosthesis became a catalyst for playfulness, with many users immediately exploring and experimenting with the potential that the augmentation allowed.

Further progress is obviously needed to address the limitations of this work, with the initial goal to be the expansion of the interactive space and to allow for mobile screens such as digital glasses that further enhance the scale and scope for the interaction. Other potential explorations would be multiple users interacting with a digital or physical object while both embodying an augmented object. Furthermore, creating augmented arms and body parts could have a further generative body change and identify potential for application in a wide range of application areas.

## 5. CONCLUSIONS

This paper has outlined initial work in the area of mixed realities, in particularly an approach for interaction with a non-real component that is not part of the fixed template of the body, or in other words the substitution of a body part with a virtual prosthesis. Whilst not fully evaluated, an initial set of experiments have shown that the approach elicits playfulness in the users of the system as they choose to interact with the digital environment in unforeseen and experimental ways.